\RequirePackage{silence}
\documentclass[fleqn,usenatbib,useAMS]{mnras} 

\usepackage{newtxtext,newtxmath}

\usepackage{graphicx}
\usepackage{amsmath}

\usepackage{amsfonts}
\usepackage{float}
\usepackage{bm}
\usepackage{ae,aecompl}
\usepackage{array}
\usepackage{soul}
\usepackage{mathtools}
\usepackage{multirow}
\usepackage[T1]{fontenc}
\usepackage[utf8]{inputenc}
\usepackage{booktabs}
\usepackage{graphicx}
\usepackage{subcaption}
\usepackage[normalem]{ulem}
\usepackage{anyfontsize}

\DeclareRobustCommand{\VAN}[3]{#2}
\let\VANthebibliography\thebibliography
\def\thebibliography{\DeclareRobustCommand{\VAN}[3]{##3}\VANthebibliography}

\DeclareRobustCommand{\appropto}{\mathrel{\vcenter{
		\offinterlineskip\halign{\hfil$##$\cr 
			\propto\cr\noalign{\kern2pt}\sim\cr\noalign{\kern-2pt}}}}}

\defcitealias{Haslbauer_2020}{HBK20}
\defcitealias{Jia_2023}{JHW23}
\defcitealias{Jia_2024}{JHW24}

\title[Redshift dependence of $H_0$ with a local void]{The redshift dependence of the inferred $H_0$ in a local void solution to the Hubble tension} 

\author[S. Mazurenko et al.]{Sergij Mazurenko$^{1}$, Indranil Banik$^{2, 3}$\thanks{E-mail: \href{mailto:sergij.mazurenko@uni-bonn.de}{sergij.mazurenko@uni-bonn.de} (Sergij Mazurenko); \newline $~~~~~~~~~~~~~~~~~\,$ \href{mailto:indranil.banik@port.ac.uk}{indranil.banik@port.ac.uk} (Indranil Banik)} and Pavel Kroupa$^{4, 5}$\vspace{10pt} \\
$^{1}$Universit\"{a}t Bonn, Regina-Pacis-Weg 3, 53115 Bonn, Germany\\
$^{2}$Scottish Universities Physics Alliance, University of Saint Andrews, North Haugh, Saint Andrews, Fife, KY16 9SS, UK\\
$^{3}$Institute of Cosmology \& Gravitation, University of Portsmouth, Dennis Sciama Building, Burnaby Road, Portsmouth PO1 3FX, UK\\
$^{4}$Helmholtz-Institut f\"{u}r Strahlen- und Kernphysik, Universit\"{a}t Bonn, Nussallee 14-16, 53115 Bonn, Germany\\
$^{5}$Astronomical Institute, Faculty of Mathematics and Physics, Charles University, V Hole\v{s}ovi\v{c}k\'ach 2, CZ-180 00 Praha 8, Czech Republic}

\date{Accepted XXX. Received YYY; in original form ZZZ}

\pubyear{\the\year{}}

\begin{document}
\label{firstpage}
\pagerange{\pageref{firstpage}--\pageref{lastpage}}
\maketitle

\begin{abstract} 
Galaxy number counts suggest that we are located within the Gpc-scale KBC void. The Hubble tension might arise due to gravitationally driven outflow from this void, as explored in detail by Haslbauer et al. We explore how the impact of the void on redshift decays at large distances. We define $H_0(z)$ as the present expansion rate $H_0$ that would be inferred from observations in a narrow redshift range centred on $z$. We find $H_0(z)$ in three different ways, all of which give similar results. We then compare these results with the observations of Jia et al., who were careful to minimize the impact of correlations between $H_0$ measurements from data in different redshift bins. We find reasonable agreement with their results for the Gaussian and Exponential void underdensity profiles, although the agreement is less good in the Maxwell-Boltzmann case. The latter profile causes severe disagreement with the observed bulk flow curve at $z < 0.1$ (Mazurenko et al.), so the tension with higher redshift data further highlights that the deepest part of the KBC void is probably near its centre. The observations show a decline of $H_0(z)$ towards the background \emph{Planck} value in qualitative agreement with the considered models, even if we use a larger void. The good overall agreement with the recent results of Jia et al. suggests that the local supervoid evident from the galaxy luminosity density out to a Gpc might also solve the Hubble tension while retaining a low background $H_0$ consistent with \emph{Planck} data, assuming enhanced structure formation on $>100$~Mpc scales.


\end{abstract}

\begin{keywords}
    cosmological parameters -- cosmology: theory -- cosmology: observations -- distance scale -- large-scale structure of Universe -- gravitation
\end{keywords}

\section{Introduction}
\label{sec:introduction}

The expansion of the Universe is parametrized by the cosmic scale factor $a$, which depends solely on time $t$ since the Big Bang and takes the value $a \equiv 1$ today. The expansion rate is given by the Hubble parameter
\begin{eqnarray}
    H ~\equiv~ \frac{\dot{a}}{a} \, ,
\end{eqnarray}
where an overdot denotes a time derivative. The functional form of $a \left( t \right)$ can be predicted using the standard model of cosmology known as the Lambda cold dark matter ($\Lambda$CDM) model \citep*{Efstathiou_1990, Ostriker_Steinhardt_1995}. It is possible to predict the present value of $H$ (denoted $H_0$) by calibrating the $\Lambda$CDM model parameters to the pattern of anisotropies in the cosmic microwave background (CMB), thought to be relic radiation from the very early Universe. CMB measurements imply that $H_0^{\mathrm{Planck}} = 67.4 \pm 0.5$~km~s$^{-1}$~Mpc$^{-1}$ \citep{Planck_2020, Tristram_2024}.

We can also obtain $H_0$ in the local Universe if we assume that it is homogeneous and isotropic on the scales used for the measurement. This appears to be the case on Gpc scales, where the radio dipole is consistent with the expectation from the Solar velocity brightening sources ahead of the Sun and dimming sources behind it \citep[][though see \citealt*{Haslbauer_2023}]{Wagenveld_2024}. With this assumption, cosmic expansion is the only source of any observed redshift
\begin{eqnarray}
    z ~\equiv~ \frac{\lambda_{\mathrm{obs}}}{\lambda_{\mathrm{emit}}} - 1 ~=~ a^{-1} - 1 \, ,
    \label{Redshift_definition}
\end{eqnarray}
where $\lambda_{\mathrm{obs}}$ is the wavelength at which we observe a feature in the spectrum of the source, $\lambda_{\mathrm{emit}}$ is the wavelength of the same feature in its rest frame, and $a$ is the cosmic scale factor when the light was emitted. Local observations can be summarized by a single number $z' \equiv dz/dr$, the slope of the redshift-distance relation in the nearby Universe. A slope arises from the fact that photons from more distant sources were emitted at an earlier time $t$. Using the chain rule, we get that
\begin{eqnarray}
    cz' ~=~ -\dot{z} ~=~ \frac{\dot{a}}{a^2} \, ,
    \label{Redshift_gradient}
\end{eqnarray}
where $c$ is the speed of light. It then follows that the local redshift gradient $cz' = \dot{a} = H_0$.

The local estimate of $H_0$ is larger than the early universe estimate based on the CMB \citep{Riess_2019, Riess_2022_comprehensive}. The anomalously high local redshift gradient of $73.0 \pm 1.0$~km~s$^{-1}$~Mpc$^{-1}$ \citep{Riess_2022_cluster} is known as the `Hubble tension' \citep[for a review, see][]{Valentino_2021}. The tension is confirmed in many different studies that use different approaches to measure extragalactic distances, some of which rely on Type Ia supernovae (SNe) while others do not \citep[][and references therein]{Scolnic_2023}. The tension is apparent in observations with the \emph{James Webb Space Telescope} (\emph{JWST}), which should be far less affected by crowding and dust extinction. \emph{JWST} observations conclusively rule out unrecognized crowding of Cepheid photometry in SN host galaxies as the cause of the Hubble tension \citep{Riess_2024_crowding}. Indeed, distances obtained by the \emph{JWST} are in good agreement with those obtained previously by the \emph{Hubble Space Telescope} \citep{Freedman_2024, Riess_2024_consistency}. The overall consistency of results obtained using a wide variety of techniques \citep{Uddin_2024} paints a clear picture that the locally measured $cz'$ exceeds the $\Lambda$CDM expectation by almost 10\% \citep[see also][]{Scolnic_2024}.

This raises the obvious possibility that the CMB anisotropies can be fit using a higher $H_0$ cosmology in which $H_0 = cz'$ within uncertainties. Such proposals require non-standard physics to ensure consistency with the observed properties of the CMB, especially the precisely measured angular scale of its first acoustic peak. Proposals along these lines are known as early time solutions to the Hubble tension. It has recently become clear that these are not viable for at least seven reasons \citep[][and references therein]{Vagnozzi_2023}. Perhaps the simplest to understand is that even if we leave aside difficulties fitting the CMB \citep{Vagnozzi_2021}, a 10\% faster expansion rate over most of cosmic history implies a 10\% younger universe. However, the ages of the oldest stars and globular clusters place a lower limit on its age of about 13.5~Gyr \citep{Cimatti_2023, Xiang_2024}. This is much more in line with $H_0^{\mathrm{Planck}}$ for plausible values of the matter density parameter $\Omega_{\mathrm{M}}$ \citep{Banik_2024}. Using differential rather than absolute ages in the cosmic chronometer (CC) technique also gives a low $H_0$ consistent with $H_0^{\mathrm{Planck}}$ but $2\sigma$ below the local $cz'$ \citep{Cogato_2024}. A subsequent analysis obtained similar results but raised the significance to $>4\sigma$ \citep{Guo_2025}. These constraints are based on stellar astrophysics, so the uncertainties affecting them are drastically different to most other cosmological analyses. For this and other reasons, it is clear that $\dot{a}$ should be close to $H_0^{\mathrm{Planck}}$, at least if we do not consider a very fine-tuned $a \left( t \right)$ which steepens suddenly at very late times \citep*{Rezazadeh_2024}.

Combining the predicted $\dot{a}$ in the \emph{Planck} cosmology with $cz'$ from the local distance ladder and adding the published fractional uncertainties in quadrature leads to the conclusion that
\begin{eqnarray}
    cz' ~=~ \left( 1.083 \pm 0.017 \right) \dot{a} \, .
    \label{cz_mismatch}
\end{eqnarray}
This directly contradicts Equation~\ref{Redshift_gradient}, implying that spectroscopic redshifts also arise for reasons beyond the change in some universal expansion factor $a \left( t \right)$ over the light travel time of a photon. This conclusion is supported by a wealth of evidence from the early and late Universe and from intermediate epochs \citep{Vagnozzi_2023}.\footnote{Redshift beyond that from cosmic expansion can arise in hybrid tired light models \citep{Kragh_2023, Gupta_2023}, but this interpretation is challenged by the observation that light from distant SNe is redshifted by the same factor as the overall light curve is stretched due to cosmological time dilation \citep{White_2024}.} Indeed, the Hubble tension is largely a tension between the local redshift gradient and estimates of the present $\dot{a}$ obtained in other ways \citep{Perivolaropoulos_2024}, with some workers even suggesting that the resolution requires ``local-scale inhomogeneous new physics disguised as local observational systematics'' \citep*{Huang_2025}.

Fortunately, $cz' \neq \dot{a}$ in general because redshifts are also caused by peculiar velocities $\bm{v}_p$, defined as velocities on top of motion that would arise in a homogeneously expanding universe. A useful working definition is that $\bm{v}_p$ is the velocity in the rest frame of the CMB. The Local Group of galaxies has $v_p = 627$~km~s$^{-1}$ \citep{Kogut_1993}, where we use the notation that $v \equiv \left| \bm{v} \right|$ for any vector $\bm{v}$. Other galaxies and clusters may have larger $v_p$. This is difficult to measure because even if the distance to an object is precisely known, the Hubble tension makes it unclear how much of its redshift should be assigned to cosmological expansion. We can instead turn to the $\Lambda$CDM theory to estimate the cosmic variance in local measurements of $cz'$, which is typically equated with $H_0$ (Equation~\ref{Redshift_gradient}). Focusing on the range $z = 0.023-0.15$ that is typically used to constrain $H_0$ locally with essentially no assumptions about the more distant Universe, the cosmic variance is only 0.9~km~s$^{-1}$~Mpc$^{-1}$ \citep{Camarena_2018}. Thus, cosmic variance cannot solve the Hubble tension if the peculiar velocities are drawn from the predicted distribution in $\Lambda$CDM \citep[c.f.][]{Wu_2017, Zhai_2022}.

\subsection{Observational evidence for a Gpc-scale local void}
\label{Local_void_evidence}

Recent work on the bulk flow of galaxies in the nearby Universe shows high peculiar velocities that are problematic in the $\Lambda$CDM framework. The bulk flow is a suitably weighted vector average of radial peculiar velocities in a spherical region centred on our location \citep{Nusser_2014, Nusser_2016}. Unlike $\bm{v}_p$ of a single object, the bulk flow within radius $r$ is not sensitive to the assumed $H_0$, making it a useful statistical probe of $\bm{v}_p$. \citet{Watkins_2023} used the CosmicFlows-4 (CF4) galaxy catalogue \citep{Tully_2023} to estimate the bulk flow on scales of $\left( 100-250 \right) h^{-1}$~Mpc, where $h \approx 0.7$ is the value of $H_0$ in units of 100~km~s$^{-1}$~Mpc$^{-1}$. \citet{Watkins_2023} found a tension of $4.81\sigma$ with the $\Lambda$CDM model $\left( P = 1.49 \times 10^{-6} \right)$ at $r = 200\,h^{-1}$~Mpc, where the bulk flow is $4\times$ the expected value. Their figure~8 shows that the tension exceeds $5\sigma$ for $r \ga 220\,h^{-1}$~Mpc. Using the same data, \citet*{Whitford_2023} independently found ``excellent agreement'' between these results and their own reported bulk flow measurements out to $173 \, h^{-1}$~Mpc, the maximum effective depth that could be probed in their more conservative approach.

These high peculiar velocities should also be associated with signs that the Universe is more structured on the relevant scales than expected. Indeed, the KBC supervoid \citep*{Keenan_2013} poses a significant problem to the $\Lambda$CDM model. It is an underdense region on a scale of around 1~Gpc \citep[see their figure~11 and figure~1 of][]{Kroupa_2015}. The void is evident at X-ray \citep{Bohringer_2015, Bohringer_2020}, optical \citep{Maddox_1990, Shanks_1990}, infrared \citep{Huang_1997, Busswell_2004, Frith_2003, Frith_2005, Frith_2006, Keenan_2013, Whitbourn_2014, Whitbourn_2016, Wong_2022}, and radio wavelengths \citep*{Rubart_2013, Rubart_2014}. Comparison with data from the Millennium XXL simulation \citep[MXXL;][]{Angulo_2012} demonstrates that the KBC void exhibits a tension of $6.04\sigma$ with the $\Lambda$CDM model (\citealt{Haslbauer_2020}, hereafter \citetalias{Haslbauer_2020}).

Various proposals have been made arguing that gravitationally driven outflows from the KBC void could solve the Hubble tension \citep{Keenan_2016, Shanks_2019a, Shanks_2019b, Ding_2020, Camarena_2022, Martin_2023}. The key idea is that since any void is underdense compared to the cosmic mean, the resulting potential hill leads to an outflow of matter away from the void, which would still be developing. To an observer situated inside the underdensity, the peculiar velocities of the surrounding matter would systematically point away, leading to an additional Doppler-redshift contribution as well as gravitational redshift (GR) from light having to climb the potential hill of the void in order to reach the observer. For a sufficiently large and deep void to occur in the $\Lambda$CDM model which would at the same time solve the Hubble tension, we would have to be located inside a $10\sigma$ primordial density perturbation \citepalias[see figure~1 of][]{Haslbauer_2020}. This led those authors to construct semi-analytic models of void formation under a force law based on Milgromian dynamics \citep[MOND;][]{Milgrom_1983, Famaey_McGaugh_2012, Banik_Zhao_2022}. MOND significantly enhances the rate of structure growth and therefore could plausibly explain the observed KBC void. \citetalias{Haslbauer_2020} developed three different void underdensity profiles (Maxwell-Boltzmann, Gaussian, and Exponential) of varying sizes and strengths in a \emph{Planck} background cosmology beginning at $z = 9$ in order to obtain a KBC-like underdensity at $z = 0$. The models were conducted in the neutrino hot dark matter ($\nu$HDM) framework \citep{Angus_2009, Katz_2013, Wittenburg_2023}. $\nu$HDM replaces the CDM component with 11~eV rest energy sterile neutrinos, allowing good fits to the internal dynamics of galaxy clusters and the CMB anisotropies \citep[see sections~7 and 9 of][respectively, and references therein]{Banik_Zhao_2022}. The models of \citetalias{Haslbauer_2020} solve the Hubble tension at late times (see their figure~6) and match the subsequently published bulk flow curve out to $z = 0.083$ \citep{Mazurenko_2024}. Accelerated structure formation compared to $\Lambda$CDM would also help to explain the formation of the El Gordo interacting galaxy clusters, whose high mass, collision velocity, and redshift are incompatible with $\Lambda$CDM at $>5\sigma$ confidence \citep*{Asencio_2021, Asencio_2023}.

An obvious consequence of a local void solution to the Hubble tension is that the effect of the void must decay at high redshift, so observations in the more distant Universe should be consistent with the \emph{Planck} cosmology. The \citetalias{Haslbauer_2020} model was tested against the observed density profile of the KBC void \citep*{Keenan_2013} and the apparent $\dot{a}$ and $\ddot{a}$ inferred from SNe out to $z = 0.15$ \citep{Camarena_2020a, Camarena_2020b}, along with a few other constraints summarized in table~4 of \citetalias{Haslbauer_2020}. Those authors also took a first exploratory step towards higher redshift tests in their figure~16, which shows how the extra redshift due to the void gradually decays away beyond the void `edge'. This was used to argue that the analysis of SNe in \citet*{Kenworthy_2019} cannot provide strong evidence against the local supervoid solution to the Hubble tension because it was assumed that $H_0(z)$ declines to $H_0^{\mathrm{Planck}}$ by $z = 0.1$, where $H_0(z)$ is the inferred value of $H_0$ from data in a narrow redshift range centred on $z$. Such a rapid decline in $H_0(z)$ is not expected in the models of \citetalias{Haslbauer_2020} as the excess redshift due to the void decays over a much larger redshift range, an issue we explore in more detail in this contribution. A subsequent analysis found that SNe show some preference for a local void when binned, but not if the unbinned data are used \citep*{Castello_2022}. We note that although parametrizing observations and theoretical predictions in terms of $H_0(z)$ misses some details especially in an inhomogeneous cosmology, it is a useful way of summarizing the data \citep[e.g.,][]{Krishnan_2020, Krishnan_2021, Cai_2021, Dainotti_2021, Dainotti_2022, Schiavone_2023, Montani_2024, Colgain_2024}.

Our main goal in this contribution is to test if the predicted decaying away of the void's effect is consistent with the latest observations. For this purpose, we define $H_0^{\mathrm{sim}}(z)$ as the $H_0$ that would be inferred by an observer at the void centre from analysing data at some redshift $z$ using Equation~\ref{Redshift_gradient}. We expect that $H_0^{\mathrm{sim}}$ converges to the background value $H_0^{\mathrm{Planck}}$ at high redshift. We consider three different techniques to obtain $H_0^{\mathrm{sim}}$ in order to bracket the range of ways in which observers might try to infer $H_0$ from data at some $z$. We then compare the predicted $H_0(z)$ with a recent compilation of observations that removes as far as possible any correlations between the results in different redshift bins, which can make it difficult to look for trends with redshift (\citealt*{Jia_2023}, hereafter \citetalias{Jia_2023}). This and other studies exploring the redshift dependence of the inferred $H_0$ were reviewed in section~2.4 of \citet{Vagnozzi_2023}. We also consider an updated version of the \citetalias{Jia_2023} analysis by \citealt*{Jia_2024} (hereafter \citetalias{Jia_2024}).

In Section~\ref{sec:methods}, we explain our three methods for obtaining $H_0^{\mathrm{sim}}$ as a function of $z$, with Method~3 being most comparable to the analysis of \citetalias{Jia_2023}. We then present our results in Section~\ref{sec:results} and discuss their implications in Section~\ref{sec:discussion}. We conclude in Section~\ref{sec:conclusions}.

\section{Methods}
\label{sec:methods}

Since our goal is to consider the predictions of the \citetalias{Haslbauer_2020} model out to cosmological distances, we need to consider the behaviour of our model in the past, when we expect peculiar velocities to have been smaller. This prevents us from using the approach adopted in \citet{Mazurenko_2024} of considering only the present-day velocity field predicted by the model. Fortunately, an important conclusion of that study was that we are located fairly close to the void centre compared to its overall size. Their table~1 indicates that the most promising scenarios put us within 200~Mpc of the void centre, which in terms of redshift is only $\approx 0.05$. The best fit to the bulk flow curve occurs if we are even closer to the void centre. Since we are interested in predictions at much larger redshifts, we will assume in what follows that we are located at the void centre. This will greatly simplify our analysis.

\begin{figure}
    \includegraphics[width=\columnwidth]{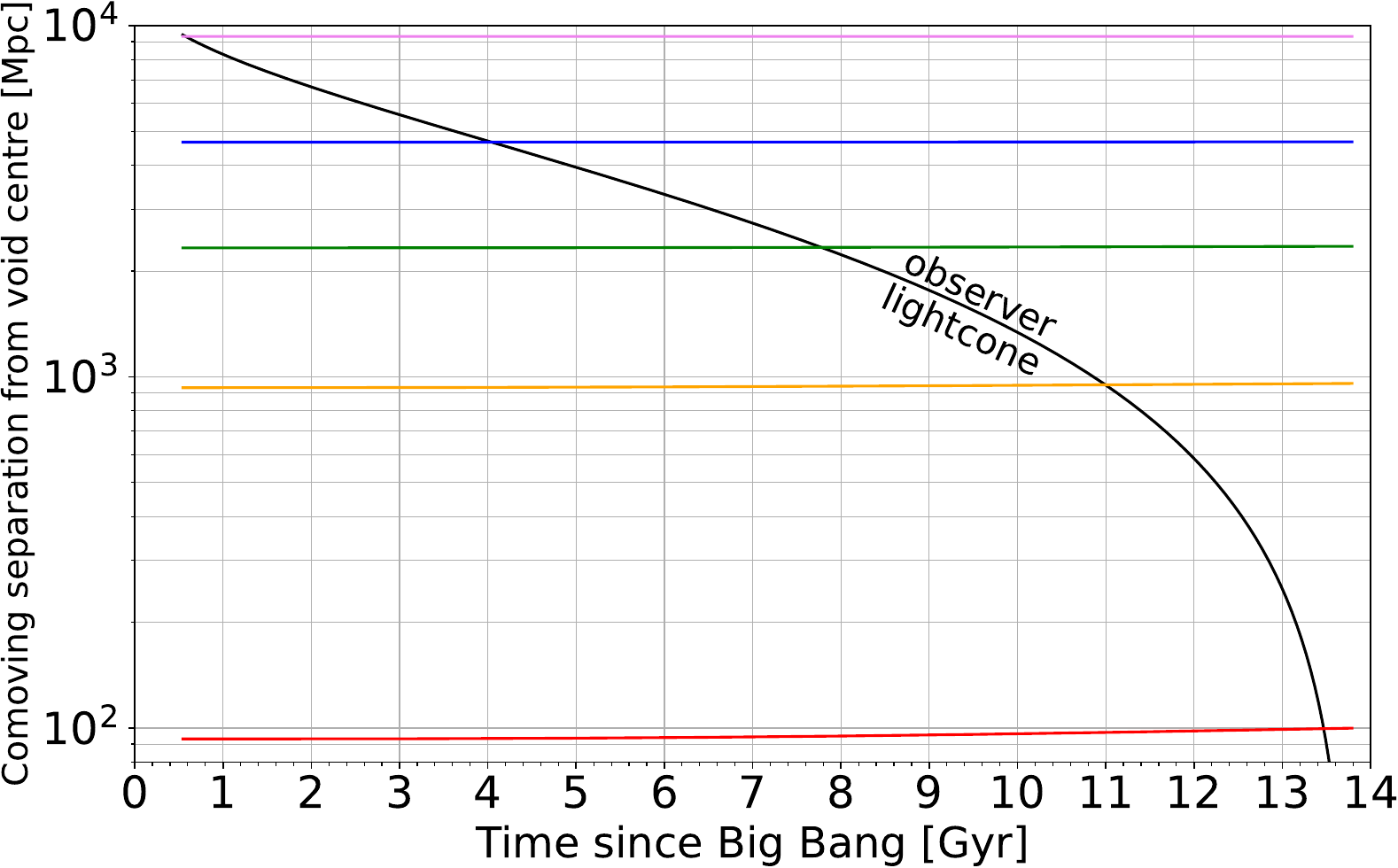}
    \caption{Illustration of how we find the intersection of our past lightcone (black; Equation~\ref{eq:Lightcone}) with the trajectories of five different particles at different initial distances from the void centre (coloured lines). Notice that particles close to the void centre have a more appreciable increase in their comoving radius due to the larger gravitational impact of the void. Results are available only for $a > 0.1$ ($z < 9$) as that is when the simulations conducted by \citetalias{Haslbauer_2020} begin.}
    \label{fig:lightcones}
\end{figure}

Our approach to handling light travel time effects follows the procedure described in section~3.3.3 of \citetalias{Haslbauer_2020}. For each of the particle trajectories in their grid, we need to find when in its history a photon emitted from the particle would by now have travelled a comoving distance equal to its comoving distance from the void centre at the time of emission. In other words, we need to find the intersection of the particle trajectory with our past lightcone. The comoving distance travelled by a photon between its emission at some time $t$ since the Big Bang and the present epoch $t_0$ is given by
\begin{eqnarray}
    r_c \left( t \right) ~=~ \int_t^{t_0} \frac{c \, dt'}{a \left( t' \right)} \, .
    \label{eq:Lightcone}
\end{eqnarray}
This is a decreasing function of $t$ that drops to 0 at the present epoch. On the other hand, the repulsive gravitational impact of a local void would cause test particles to have a rising $r_c$ curve, i.e., the physical distance $r$ from the void centre would increase faster than $a$ such that $r_c \equiv r/a$ would rise. This is illustrated in Figure~\ref{fig:lightcones}, which shows that particles at larger distances are less affected by the void and have almost constant $r_c$, as expected. The intersection of the observer lightcone with the particle trajectory determines the epoch at which we would observe the particle. We find this using the classical Newton-Raphson algorithm, making use of the peculiar velocity of the particle and the Hubble parameter to correct any mismatch in $r_c$ and ultimately reduce it to $<0.1\%$.

We then find the redshift of the particle following equation~52 of \citetalias{Haslbauer_2020}, which we reproduce for clarity.
\begin{eqnarray}
    1 + z ~=~ \frac{1}{a \left( t \right)} \overbrace{\sqrt{\frac{c + v_{\mathrm{int}}}{c - v_{\mathrm{int}}}}}^{\text{Doppler}}  \overbrace{\exp \left( \frac{1}{c^2} \int g_{\mathrm{void}} \, dr \right)}^{\text{GR}} \, ,
    \label{z_contributions}
\end{eqnarray}
where $a \left( t \right)$ is the cosmic scale factor at the time $t$ when the photon was emitted from the source, $v_{\mathrm{int}}$ is the void outflow velocity at that time and comoving distance from the void centre, and $g_{\mathrm{void}}$ is the outwards gravity from the void. We neglect the systemic velocity of the void as a whole because this would reduce the redshifts in some directions and raise it in others, so the impact of the void's overall motion is assumed to cancel out when considering observations across a wide range of directions on the sky. The contributions not from cosmic expansion have been labelled to clarify that redshifts can also arise due to the special relativistic Doppler effect and due to GR, with some approximations made when evaluating the GR contribution \citepalias[see section~3.3.1 of][]{Haslbauer_2020}. The GR term arises because photons from the particle need to climb a potential hill to reach the void centre.

Once we have the redshift $z$ of the photon, we invert Equation~\ref{Redshift_definition} to find $a_{\mathrm{app}}$, the value of $a$ at the time of emission that would be deduced by an observer from the redshift without taking into account the local void:
\begin{eqnarray}
    a_{\mathrm{app}} ~\equiv~ \frac{1}{1 + z} \, .
    \label{eq:a_app}
\end{eqnarray}
We then use $a_{\mathrm{app}}$ and the time of emission to deduce $H_0^{\mathrm{sim}}$ with three different methods, which we describe in the following sections.

Our main assumption is that observables other than redshift place constraints on how far a photon has travelled and thus on the lookback time $t_b \equiv t_0 - t$. For instance, the apparent magnitude of SNe combined with a local calibration of their absolute magnitude tells us the luminosity distance, which in the local Universe can be divided by $c$ to get $t_b$ independently of any cosmological model or the observed redshift. We therefore assume that any difference between the actual value of $H_0$ in our model (which adopts a \emph{Planck} background cosmology) and $H_0^{\mathrm{sim}}$ is entirely due to the difference
\begin{eqnarray}
    \Delta a ~\equiv~ a - a_{\mathrm{app}} \, .
    \label{Delta_a}
\end{eqnarray}
These errors in the estimated $a$ at the time of emission due to inhomogeneity will be central to our analysis.

At low redshift, it is possible to approximate $a$ and $a_{\mathrm{app}}$ as quadratic functions of time. This standard Taylor series approach is known as a cosmographic expansion, with $\dot{a} = H_0$ and $\ddot{a}$ related to the so-called deceleration parameter. To avoid unnecessary use of minus signs and fix a historical mistake in the expected sign of $\ddot{a}$, \citetalias{Haslbauer_2020} defined the acceleration parameter
\begin{eqnarray}
    \overline{q}_0 ~\equiv~ \frac{a\ddot{a}}{\dot{a}^2} \, ,
    \label{q_0_bar}
\end{eqnarray}
which is the negative of the conventionally defined deceleration parameter. Without allowing for peculiar velocities or GR, observers located within the \citetalias{Haslbauer_2020} model would estimate the cosmographic parameters by considering the run of $a_{\mathrm{app}}$ with $t_b$. The use of a Taylor series approximation to $a_{\mathrm{app}} \left( t \right)$ requires a restriction on the maximum redshift of the data that can be considered, with a typical choice being to use the range $z = 0.023 - 0.15$ \citep{Camarena_2020a, Camarena_2020b}. Their joint constraint on $H_0$ and $\overline{q}_0$ is shown in figure~6 of \citetalias{Haslbauer_2020}, alongside model predictions for all three different void density profiles that they considered. It is clear that their void models provide a reasonable match to the local observations out to $z = 0.15$. Our main goal in this contribution is to test if this is also true at higher $z$, taking advantage of subsequently published observational results while retaining the original void models of \citetalias{Haslbauer_2020}.

\subsection{Method 1}
\label{sec:Method_1}

The first and simplest method that we consider is to find the fraction by which $z$ exceeds the purely cosmological contribution from an expanding metric (Equation~\ref{Redshift_definition}):
\begin{eqnarray}
    \frac{H_0^{\mathrm{sim, 1}}}{H_0} ~=~ \frac{z}{a^{-1} - 1} \, .
    \label{eq:H0_method_1}
\end{eqnarray}
This captures the intuitive notion that if the redshift at some fixed lookback time is 5\% higher than it would be in a homogeneously expanding universe, then observers who assume it is homogeneous would overestimate the expansion rate by 5\%.

\subsection{Method 2}
\label{sec:Method_2}

Our second method is similar to our first, but with a focus on $a$ instead of $z$. Since $H_0 \equiv \dot{a}$ today, we find the time-averaged increment to $\dot{a}$ over the period between $t$ and $t_0$:
\begin{eqnarray}
    H_0^{\mathrm{sim, 2}} ~=~ H_0 + \frac{\Delta a}{t_b} \, .
    \label{eq:H0_method_2}
\end{eqnarray}
This should work very well at low $z$ because $a$ can be considered to vary linearly with time while the photon is travelling. Indeed, $a \appropto t$ over most of cosmic history \citepalias[see figure~12 of][]{Haslbauer_2020}.

\subsection{Method 3}
\label{sec:Method_3}

The third method we consider is much more complicated, but also most closely resembles the approach used by \citetalias{Jia_2023}. Instead of considering $\Delta a$, we focus on $a_{\mathrm{app}}$ itself. The idea is to find an alternative expansion history in which $a = a_{\mathrm{app}}$ at a lookback time of $t_b$. We fix the matter and dark energy density parameters \citep*[c.f.][]{Lin_2021}, but we consider a revised $H_0^{\mathrm{sim}}$. Each time this is varied, the age of the universe must be recalculated, altering $t_0$ according to equation~45 of \citetalias{Haslbauer_2020}. We then find the value of $a$ in this trial expansion history when $t = t_0 - t_b$. At that time, $a \neq a_{\mathrm{app}}$ in general, but we reduce the mismatch to a negligible level using the Newton-Raphson algorithm and record what $H_0^{\mathrm{sim}}$ was used to achieve this success.

\section{Results}
\label{sec:results}

\begin{figure}
    \includegraphics[width=\columnwidth]{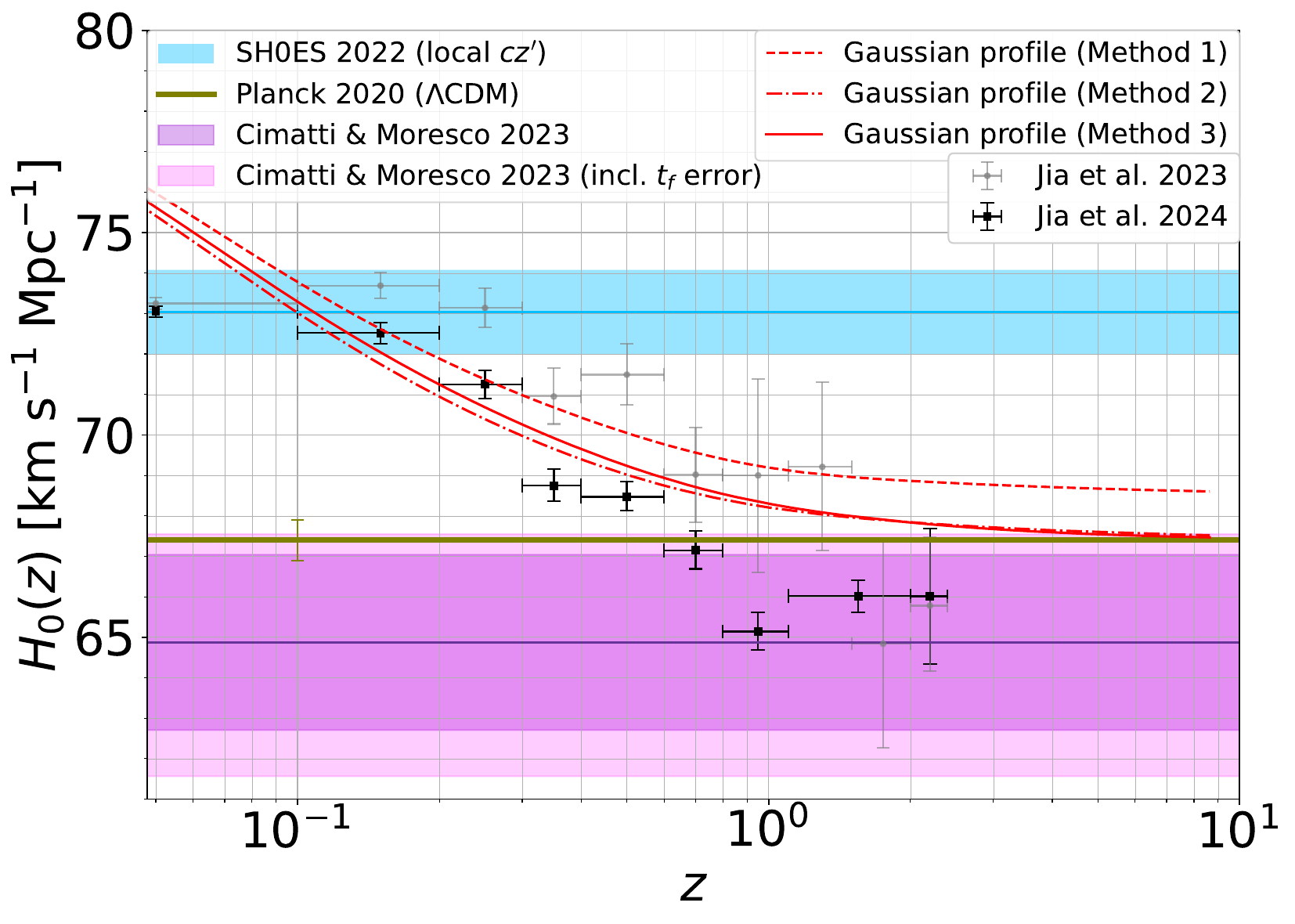}
    \caption{The red lines show the predicted $H_0(z)$ in the local void models of \citetalias{Haslbauer_2020}, as found using three different methods (Section~\ref{sec:methods}) and indicated in the legend. For clarity, results are shown here only for their best-fitting void parameters assuming a Gaussian underdensity profile. The grey points with error bars are observational results from table~4 of \citetalias{Jia_2023}, with the horizontal uncertainty being the width of each redshift bin used to infer $H_0$ independently of the data in other redshift bins. The black points are obtained similarly from an updated version of their analysis \citepalias{Jia_2024}. The solid olive line shows $H_0^{\mathrm{Planck}} = 67.4 \pm 0.5$~km~s$^{-1}$~Mpc$^{-1}$, with the uncertainty indicated using an error bar towards the left \citep{Planck_2020}. This is the background value of $\dot{a}$ today in the models shown here. The local Universe measurement \citep[$H_0^{\mathrm{SH0ES}} = 73.04 \pm 1.04$~km~s$^{-1}$~Mpc$^{-1}$;][]{Riess_2022_comprehensive} is shown as a shaded cyan band. The dark magenta band shows $H_0$ from the ages of old objects assuming they formed in $t_{\mathrm{f}} = 0.2$~Gyr and have an inverse variance weighted mean age of $t_{\mathrm{age}} = 14.05 \pm 0.25$~Gyr \citep{Cimatti_2023}. The lighter magenta band also includes a factor of 2 uncertainty in $t_{\mathrm{f}}$. We obtain $H_0$ from the cosmic age $t_0 = t_{\mathrm{age}} + t_{\mathrm{f}}$ using equation~45 of \citetalias{Haslbauer_2020} assuming $\Omega_{\mathrm{M}} = 0.302$ \citep*{Lin_2021}, neglecting its uncertainty of 0.008 \citep[see also][]{Banik_2024}.}
    \label{fig:H_z_over_z_Gaussian_all_methods}
\end{figure}

Using the methods described in Section~\ref{sec:methods}, we use Figure~\ref{fig:H_z_over_z_Gaussian_all_methods} to plot how the apparent $H_0$ should depend on the redshift of the data used to obtain it. The observational results are from \citetalias{Jia_2023}, while the model predictions are limited to the Gaussian density profile because this provides a good match to the observed bulk flow curve \citep{Mazurenko_2024} $-$ other density profiles will be considered shortly. As might be expected, all three methods give a high local value of $H_0$, but the $H_0(z)$ curve then declines towards $H_0^{\mathrm{Planck}}$ at high $z$. The convergence to the \emph{Planck} cosmology is not complete with Method~1 because the GR contribution arises mostly from the late stages of the photon's journey to our detectors. This causes a fixed percentage increase to the redshift, which in Method~1 causes a fixed percentage overestimation of $H_0$ (Equation~\ref{eq:H0_method_1}). This does not arise with the other methods because they focus on $a$, which becomes small at high redshift. Thus, the impact of the GR contribution to $a_{\mathrm{app}}$ also becomes small at high redshift, even though this is not true in fractional terms. As a result, Methods~2 and 3 converge more rapidly to $H_0^{\mathrm{Planck}}$, which is the actual value of the present $\dot{a}$ in all models.

Moving forwards, our focus will be on the results obtained using Method~3 as this is the method which is most similar to the approach used by \citetalias{Jia_2023} and \citetalias{Jia_2024}. Results obtained via Methods 1 and 2 are similar to those using Method~3 and can be found in Appendix~\ref{sec:curves_method_1_2}.

\begin{figure*}
    \includegraphics[width=\textwidth]{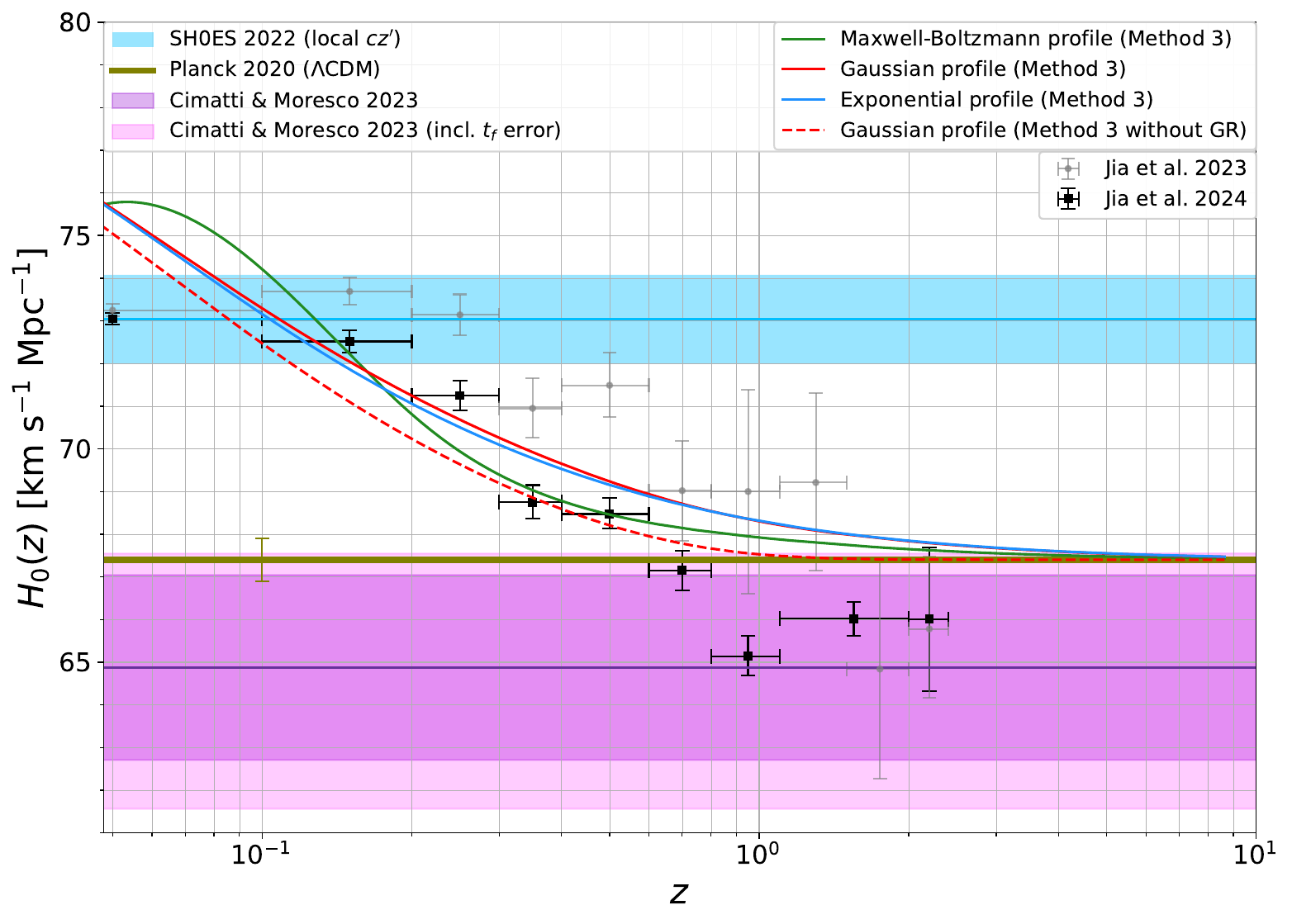}
    \caption{Similar to Figure~\ref{fig:H_z_over_z_Gaussian_all_methods}, but now showing $H_0(z)$ obtained via Method~3 (Section~\ref{sec:Method_3}) for the Maxwell-Boltzmann, Gaussian, and Exponential density profiles (solid green, red, and blue curves, respectively). The dashed red curve shows results for the Gaussian density profile if we neglect the GR term in Equation~\ref{z_contributions}, helping to show its impact.}
    \label{fig:Hubble_Diagram_Method_3_Jia_plus_noGR}
\end{figure*}

Our main result is Figure~\ref{fig:Hubble_Diagram_Method_3_Jia_plus_noGR}, which shows the results from Method~3 for all three void density profiles considered by \citetalias{Haslbauer_2020}. The void parameters used for each profile are the best-fitting values as published in that study. The data from \citetalias{Jia_2023} and \citetalias{Jia_2024} are shown as points that are plotted at the centre of each redshift bin used, with the plotted horizontal uncertainties showing the width of each bin. The $H_0(z)$ curves for the Maxwell-Boltzmann, Gaussian, and Exponential void density profiles are shown in solid green, red, and blue, respectively. The model-dependent early universe measurement of $H_0$ using the CMB is shown as a solid olive line \citep[$H_0^{\mathrm{Planck}} = 67.4 \pm 0.5$~km~s$^{-1}$~Mpc$^{-1}$;][]{Planck_2020}, while the late or local Universe measurement of $cz'$ \citep[$H_0^{\mathrm{SH0ES}} = 73.04 \pm 1.04$~km~s$^{-1}$~Mpc$^{-1}$;][]{Riess_2022_comprehensive} is shown as a shaded cyan band under the usual assumption that $H_0 = cz'$ (Equation~\ref{Redshift_gradient}). The constraint on $H_0$ from the ages of old stars and stellar populations is shown using magenta bands on Figure~\ref{fig:Hubble_Diagram_Method_3_Jia_plus_noGR}, with the more darkly shaded region corresponding to the formation time being $t_{\mathrm{f}} = 0.2$~Gyr and the less darkly shaded region allowing $t_{\mathrm{f}}$ to be uncertain by a factor of 2 \citep{Cimatti_2023}. Both this age constraint and the results of \citetalias{Jia_2024} at the highest probed redshifts least affected by any local void suggest that $\dot{a}$ is $1-2$~km~s$^{-1}$~Mpc$^{-1}$ lower than $H_0^{\mathrm{Planck}}$ \citep[see also][]{Farren_2024, Ge_2024}.

To help disentangle the different non-cosmological contributions to the redshift (Equation~\ref{z_contributions}), we use a dashed red curve in Figure~\ref{fig:Hubble_Diagram_Method_3_Jia_plus_noGR} to show results for the Gaussian void profile without including the GR term. Since any deviation between $H_0^{\mathrm{sim}}$ and $H_0$ arises entirely due to the non-cosmological contributions from GR and the special relativistic Doppler effect from peculiar velocities, the dashed red curve shows the impact of considering only the latter, while the solid red curve considers both. It is apparent that GR boosts $H_0^{\mathrm{sim}}$ by $\la 1$~km~s$^{-1}$~Mpc$^{-1}$ at all redshifts, with the effect being maximal at $z \approx 0.5$. At higher redshift, the GR contribution begins to overtake the redshift from the special relativistic Doppler effect. This is because the height of our local potential hill creates an extra contribution to the redshift that affects photons coming from the distant universe, where any outflow induced by a local void would be very small. The GR term would also induce a slight redshift of the surface of last scattering, which would slightly affect the temperature of the CMB at the present epoch. However, this effect on the CMB monopole is very small and well within current uncertainties \citepalias[see section~5.3.3 of][]{Haslbauer_2020}.

It is important to emphasize here that our $H_0(z)$ results should not be over-interpreted at low redshift because of our assumption that we are located at the void centre (Section~\ref{sec:methods}). However, the bulk flow observations of \citet{Watkins_2023} imply a slightly off-centre location roughly $100 - 150$~Mpc from the void centre \citep{Mazurenko_2024}. For this reason, all our figures showing $H_0(z)$ curves are restricted to $z \geq 0.05$.

\section{Discussion}
\label{sec:discussion}


Our analysis is based on the local supervoid model of \citetalias{Haslbauer_2020}, which was designed with a background \emph{Planck} cosmology that has $H_0 = H_0^{\mathrm{Planck}} = 67.4 \pm 0.5$~km~s$^{-1}$~Mpc$^{-1}$ \citep{Planck_2020}. The \citetalias{Haslbauer_2020} models were constrained to fit the density profile of the KBC void \citep*{Keenan_2013} and use outflows from it to also fit the locally measured $H_0$ and $\overline{q}_0$ (Equation~\ref{q_0_bar}). Remarkably, the same void model later turned out to fit the rising galaxy bulk flow curve observed by \citet{Watkins_2023}, even though the bulk flows on the largest probed scales are about $4\times$ the $\Lambda$CDM expectation and thus disagree with it at $>5\sigma$ confidence.

Since the \citetalias{Haslbauer_2020} models assume a local void with a background \emph{Planck} cosmology, it is not surprising that the simulated $H_0(z)$ curves all show a transition from the high local $cz'$ down to the low \emph{Planck} value at high redshift. When comparing the detailed form of the predicted transition to observations of this transition, it is important to note that the Maxwell-Boltzmann density profile has been ruled out as a possible candidate for the density distribution in the local universe \citep{Mazurenko_2024} due to incompatibility with the observed rising bulk flow curve \citep{Watkins_2023, Whitford_2023}. The Maxwell-Boltzmann profile is interesting because it is the only profile we consider in which the void is not deepest at its centre. While this may seem somewhat contrived and is not in line with the bulk flow measurements, we nevertheless show results using this profile in order to demonstrate the impact of the density profile on the $H_0(z)$ curve. It is beneficial to have the complete set of such curves for all three void density profiles considered by \citetalias{Haslbauer_2020}, especially when the Gaussian and Exponential profiles give very similar results (compare the solid red and blue curves in Figure~\ref{fig:Hubble_Diagram_Method_3_Jia_plus_noGR}).

The Maxwell-Boltzmann profile has a much faster convergence to $H_0^{\mathrm{Planck}}$ because the void is much smaller. This could be due to observations not supporting the lack of any appreciable underdensity in the central regions of the void, forcing the \citetalias{Haslbauer_2020} analysis to prefer models which limit the size of this region. However, this also makes the whole void smaller, causing a faster decay of $\Delta a$ (Equation~\ref{Delta_a}) and thus $H_0^{\mathrm{sim}} - H_0$. In particular, $H_0^{\mathrm{sim}}$ becomes indistinguishable at $1\sigma$ confidence from the underlying $H_0 = H_0^{\mathrm{Planck}}$ by the time we get to $z \approx 1$ (green curve in Figure~\ref{fig:Hubble_Diagram_Method_3_Jia_plus_noGR}). However, with the Gaussian (solid red) and Exponential (blue) density profiles, $H_0^{\mathrm{sim}}$ gets within $1\sigma$ of $H_0^{\mathrm{Planck}}$ only when $z \ga 1.8$. As a result, the Maxwell-Boltzmann profile causes some tension with the observations, especially at $z \approx 0.5$. This is completely independent of the tensions it faces with the bulk flow curve at $z < 0.1$ \citep{Mazurenko_2024}.\footnote{The velocity fields predicted by all three profiles are presented in its appendix.} However, it is important to note that the more recent results of \citetalias{Jia_2024} are more in line with the Maxwell-Boltzmann profile because this updated study indicates that $H_0(z)$ converges to $H_0^{\mathrm{Planck}}$ much more rapidly.



Remarkably, the $H_0(z)$ curves for the Gaussian and Exponential density profiles are almost identical across the entire redshift range. The high degree of similarity arises because both profiles place the deepest part of the void at its centre, which provides a much better match to the bulk flow curve. Near the void centre, the velocity fields and height of the potential hill are similar enough that the combined contribution from peculiar velocities and GR (Equation~\ref{z_contributions}) result in approximately the same contribution to the total redshift. This is to be expected because the three investigated best-fitting models from \citetalias{Haslbauer_2020} were shown to solve the Hubble tension (see their figure~6), so the non-cosmological contributions to the redshift will naturally be similar in the local universe.

At high redshift, any observed object is no longer part of the void, so the local density near the object is close to the cosmic mean density. The Doppler redshift contribution from peculiar velocities (Equation~\ref{z_contributions}) becomes negligible compared to the redshift from cosmic expansion. However, because the object lies beyond the extent of the void, light originating from the object has to climb the entirety of the void's potential hill. This leads to a similar GR contribution for all three density profiles. The effect of GR is highlighted in Figure~\ref{fig:Hubble_Diagram_Method_3_Jia_plus_noGR} using the solid and dashed red lines, which show the $H_0(z)$ curve for the Gaussian density profile with and without the GR contribution, respectively. At $z \la 0.1$, GR has only a small effect of $\la 1$~km~s$^{-1}$~Mpc$^{-1}$ because photons have to climb less of the void's potential hill, while peculiar velocities are rather high in this region. GR starts to dominate over the Doppler term as the main non-cosmological source of redshift by $z \ga 0.5$.


Our results demonstrate that as we go out to larger redshift in the \citetalias{Haslbauer_2020} local void model, $H_0^{\mathrm{sim}}(z)$ approaches the global value $H_0^{\mathrm{Planck}}$ in a way that broadly agrees with the analysis of \citetalias{Jia_2023} for the Gaussian and Exponential density profiles. At intermediate $z$, there are some differences between the observed $H_0(z)$ curve and the predictions with these two profiles, both of which somewhat underestimate the empirical analysis of \citetalias{Jia_2023}. Before discussing possible reasons for this tension in the next section, it is worth reflecting on the more general point that a declining trend in the observed $H_0(z)$ strongly argues against pre-recombination solutions to the Hubble tension \citep[as also argued in section~2.4 of][]{Vagnozzi_2023}. One may worry that the declining trend observed by \citetalias{Jia_2023} is caused by a transition from the data being dominated by SNe at low $z$ to baryon acoustic oscillations (BAOs) at high $z$, given these techniques are most accurate in different redshift ranges. Their BAO results assume that the BAO ruler has a size of 147.5 comoving Mpc (cMpc) based on assuming standard pre-recombination physics. This can have the effect of creating a circular argument in which BAO results are guaranteed to return $H_0^{\mathrm{Planck}}$ even if the actual comoving size of the standard ruler differs from the assumed value, as long as it has stayed the same when $z \la 10$. However, figure~2 of \citetalias{Jia_2023} shows that SNe alone give a declining $H_0(z)$ trend, albeit with larger uncertainties. The highest redshift data point has $H_0(z)$ almost exactly equal to $H_0^{\mathrm{Planck}}$. Moreover, time delay cosmography of the multiply lensed SN Refsdal gives $H_0 = 64.8^{+4.4}_{-4.3}$~km~s$^{-1}$~Mpc$^{-1}$ in a blinded analysis \citep{Kelly_2023}, almost $2\sigma$ below the local $cz'$. The lens is at $z = 0.54$ and the SN itself is at $z = 1.49$, so our results suggest that we should expect $H_0^{\mathrm{sim}} = 68-69$~km~s$^{-1}$~Mpc$^{-1}$ depending on which is most relevant (Figure~\ref{fig:Hubble_Diagram_Method_3_Jia_plus_noGR}). Interestingly, the lensing result is based on 8 cluster lens models, but restricting to the two which best match the observations gives $H_0 = 66.6^{+4.1}_{-3.3}$~km~s$^{-1}$~Mpc$^{-1}$, which is even closer to $H_0^{\mathrm{Planck}}$. If the Hubble tension were caused by systematics in the local measurement of $cz'$ (which we argued against in Section~\ref{sec:introduction}), it would be a remarkable coincidence for $H_0(z)$ inferred from distant SNe to match $H_0^{\mathrm{Planck}}$ given that SNe are observed and analysed completely independently of the CMB. It would be even more surprising if the combination of SN, BAO, and CC data analysed by \citetalias{Jia_2023} and \citetalias{Jia_2024} give an observationally inferred $H_0(z)$ curve that roughly matches the expectation of the previously published \citetalias{Haslbauer_2020} model. Our results therefore seem to support their local void scenario for the Hubble tension. But the agreement with the \citetalias{Jia_2023} observations is not perfect at intermediate $z \approx 0.1 - 0.6$, an issue we turn to next. We note that the model predictions fall roughly between the observational results of \citetalias{Jia_2023} and \citetalias{Jia_2024}.


\subsection{Comparison with larger void sizes}
\label{sec:discussion larger void}

The observed $H_0(z)$ curve reported by \citetalias{Jia_2023} declines somewhat more gradually than expected in the \citetalias{Haslbauer_2020} model at $z \la 0.6$, even if we neglect the Maxwell-Boltzmann profile that is also problematic for other reasons. One possible interpretation is that the void is larger than assumed in their study. Those authors used a prior in which the initial ($z = 9$) void size was limited to 1030~cMpc (see their table~1). This turned out to be rather restrictive for the Gaussian and Exponential profiles, since in both cases the posterior on the void size rises rapidly and does not peak before the maximum allowed value (see their appendix~C). This suggests that if a wider prior had been considered, the overall fit to the data would improve. Note this is based purely on the data available at that time: obviously this preference for large voids is unrelated to the subsequent results of \citetalias{Jia_2023}. It might instead be related to the somewhat high $\overline{q}_0$ in these models, or the inclusion of strong lensing time delays.

\begin{figure}
    \includegraphics[width=\columnwidth]{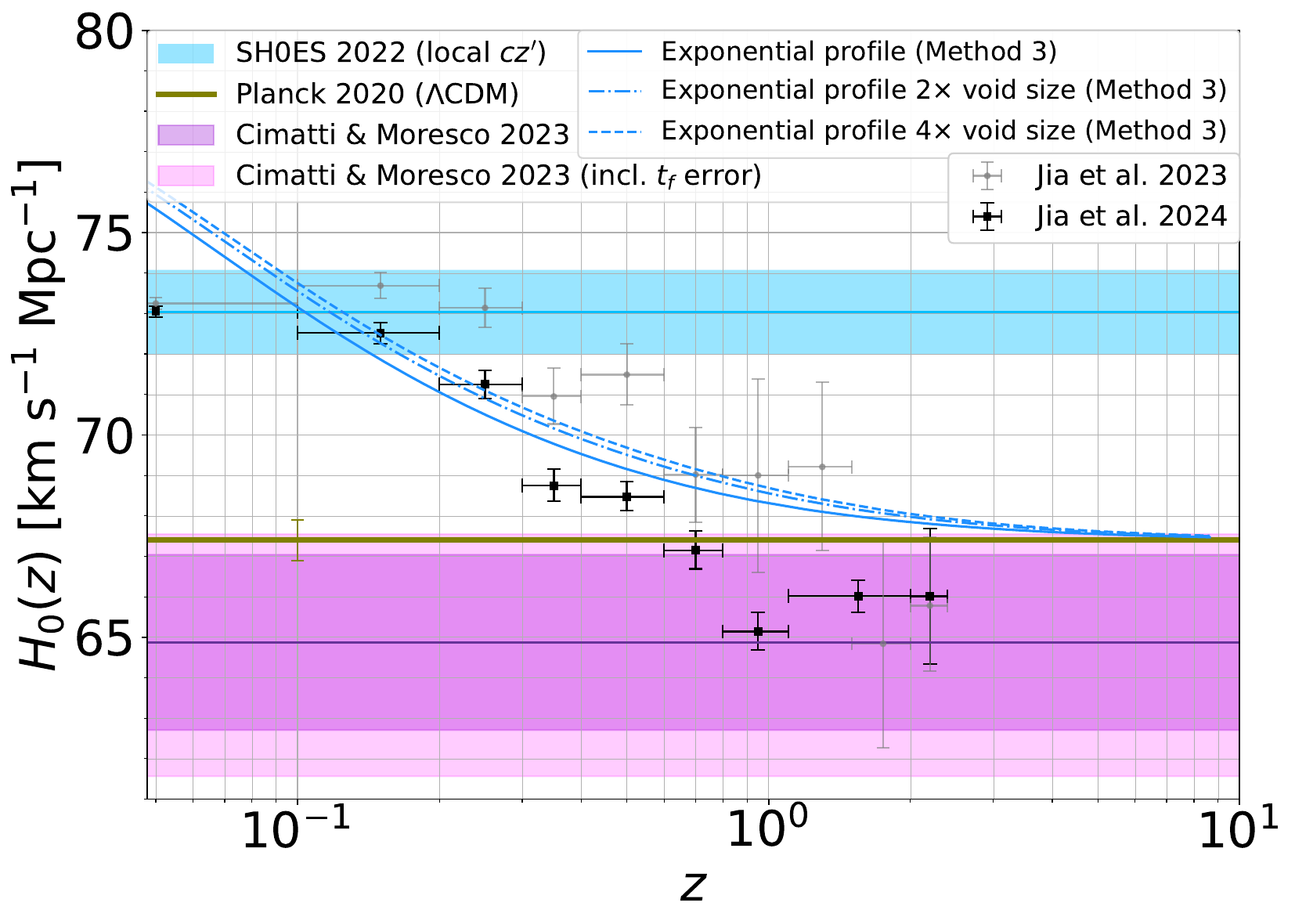}
    \caption{Similar to Figure~\ref{fig:Hubble_Diagram_Method_3_Jia_plus_noGR}, but only for the Exponential density profile. The $H_0(z)$ curves are shown for a void with the original, doubled, and quadrupled initial comoving void size using the solid, dashed, and dash-dotted blue line, respectively.}
    \label{fig:Hubble_Diagram_voidsize_variation}
\end{figure}

To explore the possibility of a larger void, we consider altering the initial void size. Results are shown in Figure~\ref{fig:Hubble_Diagram_voidsize_variation}, focusing on the Exponential density profile. We show the predicted $H_0(z)$ with the original void size (solid blue line) alongside results with a $2\times$ larger void (dashed blue line) and a $4\times$ larger void (dash-dotted blue line). All other void parameters remain the same in all three cases. Our results show that the $H_0(z)$ curve is rather insensitive to the void size. Although adopting a larger void does boost the $H_0(z)$ curve upwards in a minor way and this does improve the agreement with observations, the results closely follow the shape of the original $H_0(z)$ curve. Interestingly, the impact of going from double to quadruple the original void size seems to be only about half the impact of doubling the original void size. These diminishing returns from enlarging the void size indicate that this cannot by itself yield full agreement with the observations of \citetalias{Jia_2023}, though interestingly the Exponential and Gaussian model predictions are approximately half-way between their results and those of \citetalias{Jia_2024}.

In the real Universe, there would also be structures besides any local void, which may affect observations at higher $z$. For instance, the apparent peak in the cosmic star formation history could be interpreted as due to a major overdensity of matter at $z \approx 1.9$ on a scale of 5~cGpc \citep*{Haslbauer_2023}. Spectroscopic studies of quasars also reveal unexpectedly large arcs and rings of absorbers along the line of sight on a scale of about 1~cGpc at $z = 0.8$ \citep*{Lopez_2022, Lopez_2024}. In the future, cosmological models allowing for such large-scale inhomogeneities need to be studied. If the observations at some $z$ lack complete sky coverage but rather focus on particular sky directions, then the combination of such structures with survey selection effects may lead to additional stochasticity in the observationally determined $H_0(z)$.

\subsection{Influence of the observer position and void geometry}
\label{sec:discussion observer position and void geometry}

Our $H_0(z)$ results are restricted to $z \geq 0.05$ because of our simplifying assumption that we have a central location in the KBC void (Section~\ref{sec:methods}). Even if we were to relax this assumption \citep[as done in][]{Mazurenko_2024}, those authors pointed out that small scale systematic effects on the velocity field due to individual galaxies and clusters inside the void cannot be accurately accounted for given the limited spatial resolution of the model. This is an aspect that was intentionally not considered by \citetalias{Haslbauer_2020} as their focus was on the large-scale structure of the void.

The position of the observer within the void should play some role in how the $H_0(z)$ curve behaves, especially at low redshift. An off-centre location $\approx 100-150$~Mpc from the void centre \citep[as suggested in][]{Mazurenko_2024} would lead to anisotropy in the apparent expansion rate depending on the line of sight of the observer. This would be an observable effect at low redshift. Interestingly, recent observational results do suggest some anisotropy in the $cz'$ inferred from SNe \citep{Javanmardi_2015, Kalbouneh_2023, Hu_2024, Sah_2025}. This is also apparent with galaxy data, though in that case the anisotropy is typically interpreted as a bulk flow \citep{Watkins_2023}. An accurate quantitative description of the anisotropy that an off-centre observer would see in the \citetalias{Haslbauer_2020} model will give a good insight into the validity of this scenario (Banik et al., in preparation).

Anisotropy would also be present at intermediate redshifts, where the discrepancy in Figure~\ref{fig:Hubble_Diagram_Method_3_Jia_plus_noGR} is most pronounced. However, the anisotropy would become smaller with increasing redshift and distance, which makes our offset from the void centre proportionally smaller. Moreover, the results of \citet{Mazurenko_2024} indicate that our offset from the void centre is small compared to the total size of the void. A more distant observed object will be located towards the edge of the void or already beyond it, making the GR contribution more important (compare the solid and dashed red lines in Figure~\ref{fig:Hubble_Diagram_Method_3_Jia_plus_noGR}). Since the potential at our location is fixed regardless of the line of sight and since a distant object would have little potential energy due to a local void, the GR contribution to the redshift would be nearly isotropic. Thus, $H_0(z)$ should not depend much on the line of sight.

An important caveat is that we have not considered the impact of the whole void's systemic velocity in the CMB frame. This is critical to include when assessing the predicted bulk flow \citep{Mazurenko_2024}, but as discussed in Section~\ref{sec:methods}, we assume its effect cancels out when combining observations from many different sky directions. However, the void's systemic velocity must be caused by even more distant structures. This may be responsible for the modest upturn in the observed $H_0(z)$ curve at $z = 0.5$ in the analysis of \citetalias{Jia_2023} \citepalias[though the feature is less pronounced in][]{Jia_2024}. Since $z = 0.5$ is quite distant for SN observations, their sky coverage could be limited at that redshift. Thus, the observed value of $H_0(z)$ may depend on which particular sky directions contribute most to the observations.

It is important to remember that the local void model of \citetalias{Haslbauer_2020} is rather simple, consisting of spherically symmetric outflow plus a systemic velocity in the CMB frame. For simplicity, the void was assumed to be spherically symmetric. The model also by design excluded any consideration of structures orders of magnitude smaller than the void itself. This was primarily done in order to limit the necessary computational resources associated with the simulation of a high number of different model permutations while maintaining adequate resolution. These simplifying assumptions will of course not hold in the real Universe. In particular, one can easily imagine that the void's isodensity contours are not spherical but have an ellipsoidal or triaxial shape. To understand the impact that this could have, we can consider an idealized scenario in which the universe was homogeneous initially apart from an oblate ellipsoidal region with a uniformly lower density. The outward gravity would be stronger along the short axis than the long axes, much like the terrestrial gravity is stronger near the poles. This behaviour is similar to the gravity from overdensities, the only difference being that the gravity from a void is outwards. Thus, overdensities would tend to become more anisotropic over time and may eventually form sheets and filaments, while underdensities would tend to become more spherical. This suggests that the approximation of a spherical void may be reasonably accurate. However, a void in the real Universe would not exist in isolation. Surrounding matter would create structures whose gravity would introduce irregularities into any void isodensity contour, be it spheroidal or ellipsoidal \citep{Wittenburg_2023}. A void finding analysis applied to $\nu$HDM cosmological simulations is currently underway to explore some of these issues (Russell et al., in preparation). In principle, anisotropy of the void's isodensity contours combined with our off-centre location within the void could modify the $H_0(z)$ curve, though further analysis of this aspect is beyond the scope of this work.

\section{Conclusions}
\label{sec:conclusions}

The gradient of redshift with respect to distance in the local Universe is steeper than expected in the $\Lambda$CDM paradigm calibrated to the pattern of anisotropies in the CMB (Equation~\ref{cz_mismatch}). This Hubble tension might be due to outflow from the observed KBC supervoid \citep*{Keenan_2013} inflating redshifts in the local Universe. This scenario was explored in detail by \citetalias{Haslbauer_2020}, who used semi-analytic models to study three different void density profiles in a background \emph{Planck} cosmology. The voids were evolved with the MOND force law, which provides the necessary significant enhancement to structure growth on $\ga 100$~Mpc scales compared to $\Lambda$CDM, this being incompatible with the observed properties of the KBC void at $6.0\sigma$. Those authors obtained best-fitting parameters in each case by comparing the results of their semi-analytic calculations to a variety of observations.

A strong prediction of a purely local KBC void-based solution to the Hubble tension is that observations which probe beyond the void would not show the Hubble tension, instead becoming consistent with the assumed \emph{Planck} background cosmology. We test this by calculating what value of $H_0$ would be inferred by observers at the centre of the \citetalias{Haslbauer_2020} void model if they only consider data in some narrow redshift range and assume homogeneity. We use three different methods to compute the $H_0(z)$ curve (Section~\ref{sec:methods}) and compare to the observational results of \citetalias{Jia_2023} and \citetalias{Jia_2024}, who follow an approach that most closely resembles our Method~3. Their results clearly show a declining $H_0(z)$, with the high redshift data coming very close to $H_0^{\mathrm{Planck}}$ \citepalias{Jia_2023} or even slightly below it \citepalias{Jia_2024}, which interestingly is more consistent with stellar ages. This already shows the promise of a late-time or local solution to the Hubble tension.

To obtain model predictions at high redshift, we need to consider lookback time effects, preventing us from using only the present velocity field of the void \citep[as done by][]{Mazurenko_2024}. To limit the complexity of our analysis, we assume that we are located at the void centre, which should be a reasonable assumption given their result that we should be located only $100-150$~Mpc away. Although the methods we use to obtain the predicted $H_0(z)$ curve are approximations and more detailed work is needed to directly compare with the underlying observables, we find reasonable agreement with the observed $H_0(z)$ curve for the Gaussian and Exponential void density profiles. There is some tension with the Maxwell-Boltzmann profile, which in any case was previously shown to predict a bulk flow curve at $z < 0.1$ that is strongly incompatible with observations \citep{Mazurenko_2024}. The observational results of \citetalias{Jia_2023} and \citetalias{Jia_2024} approximately straddle the model predictions (Figure~\ref{fig:Hubble_Diagram_Method_3_Jia_plus_noGR}), which are not much affected purely by enlarging the void size (Figure~\ref{fig:Hubble_Diagram_voidsize_variation}).

It should be borne in mind that the \citetalias{Haslbauer_2020} model is very simple, consisting only of spherically symmetric outflow combined with a systematic velocity of the whole void. While isolated voids should become more spherical with time (opposite to overdensities), the real Universe appears to have a more varied and larger scale inhomogeneous structure \citep*{Haslbauer_2023}, complicating the picture. It is possible that the predicted $H_0(z)$ would be boosted at intermediate redshift in a detailed analysis of a more complex void geometry influenced by surrounding matter, combined with a better accounting of our location within the void and the impact of incomplete sky coverage in some of the measurements used to constrain $H_0(z)$. This will require Gpc-scale self-consistent cosmological $\nu$HDM simulations (Russell et al., in preparation). The analysis should include insights from prior work such as the bulk flow \citep{Mazurenko_2024} and slightly smaller scale hydrodynamical cosmological $\nu$HDM simulations \citep{Wittenburg_2023}.

Our results suggest that a local underdensity is a viable solution to the Hubble tension which not only matches constraints from the CMB and the $z < 0.15$ Universe \citepalias[figure 6 of][]{Haslbauer_2020}, but also broadly agrees with constraints from intermediate redshifts. Moreover, galaxy number counts continue to show a local underdensity \citep{Wong_2022} and it remains difficult to solve the Hubble tension by modifying pre-recombination physics \citep{Philcox_2022, Vagnozzi_2023, Toda_2024, Zaborowski_2024}. We therefore plan to further explore the local void scenario by obtaining predictions for the actual observables, especially those related to BAOs (Banik \& Kalaitzidis, in preparation). These may hold important information because the BAO signal parallel to the line of sight and within the plane of the sky would in general be affected differently by a local void, a subtlety not considered here. It is also important to consider SNe at both high redshift and in the local Universe, where the redshift gradient would inevitably be somewhat anisotropic in the local void scenario.

\section*{Acknowledgements}

IB is supported by Royal Society University Research Fellowship grant 211046 and was supported by Science and Technology Facilities Council grant ST/V000861/1. PK thanks the Deutscher Akademischer Austauschdienst-Eastern European exchange programme for support. IB thanks Vasileios Kalaitzidis for calculating $H_0$ from the age of the Universe, a constraint shown in most of our figures. The authors are grateful to the referees for comments that significantly improved this contribution.

\section*{Data Availability}

The simulation results were previously published in \citetalias{Haslbauer_2020} along with the parameters of the most likely model for each void density profile.

\bibliographystyle{mnras}
\bibliography{Hubble_diagram_Jia}

\begin{appendix}

\section{Results using Methods 1 and 2}
\label{sec:curves_method_1_2}

\begin{figure*}
    \includegraphics[width=\columnwidth]{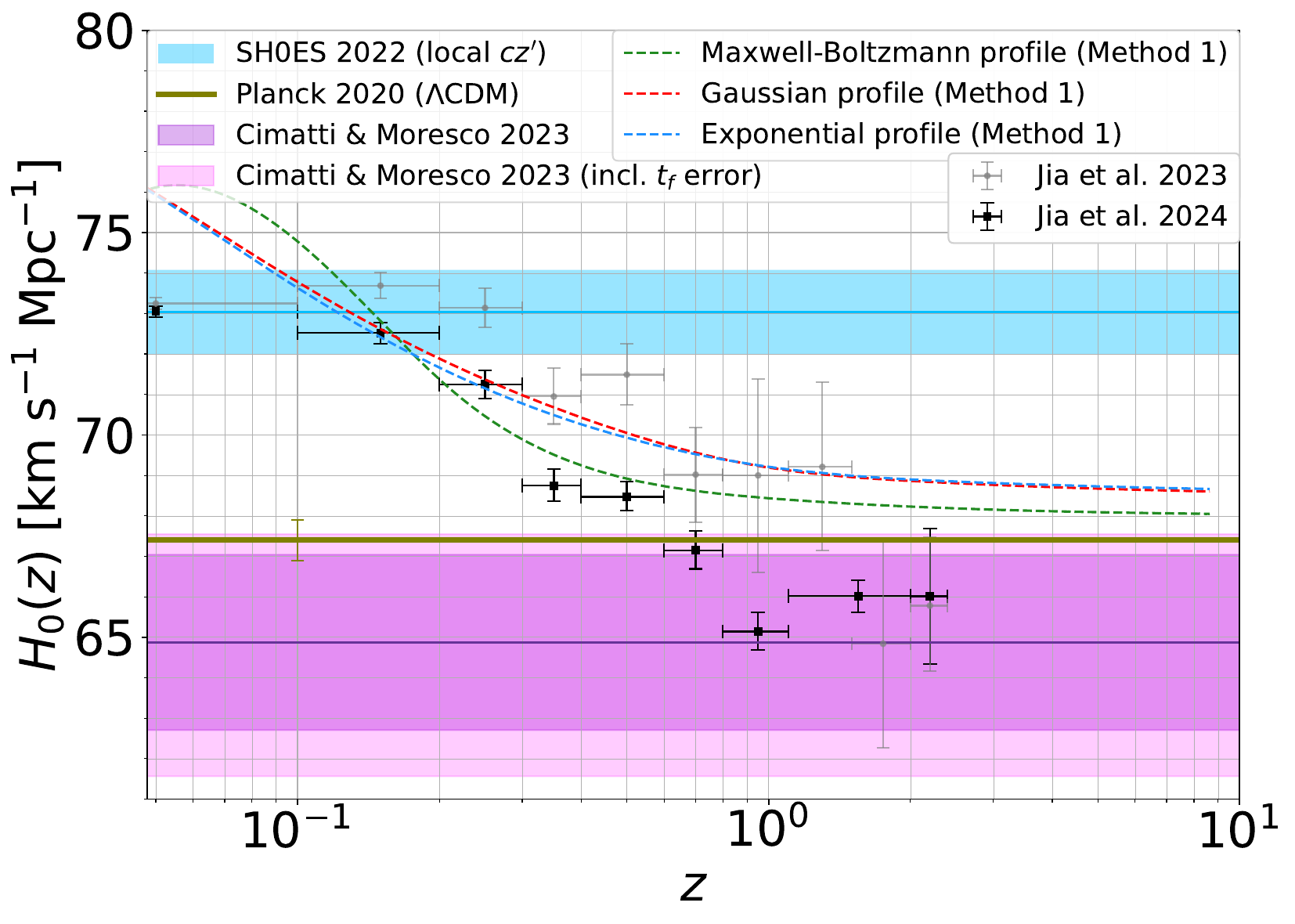}
    \includegraphics[width=\columnwidth]{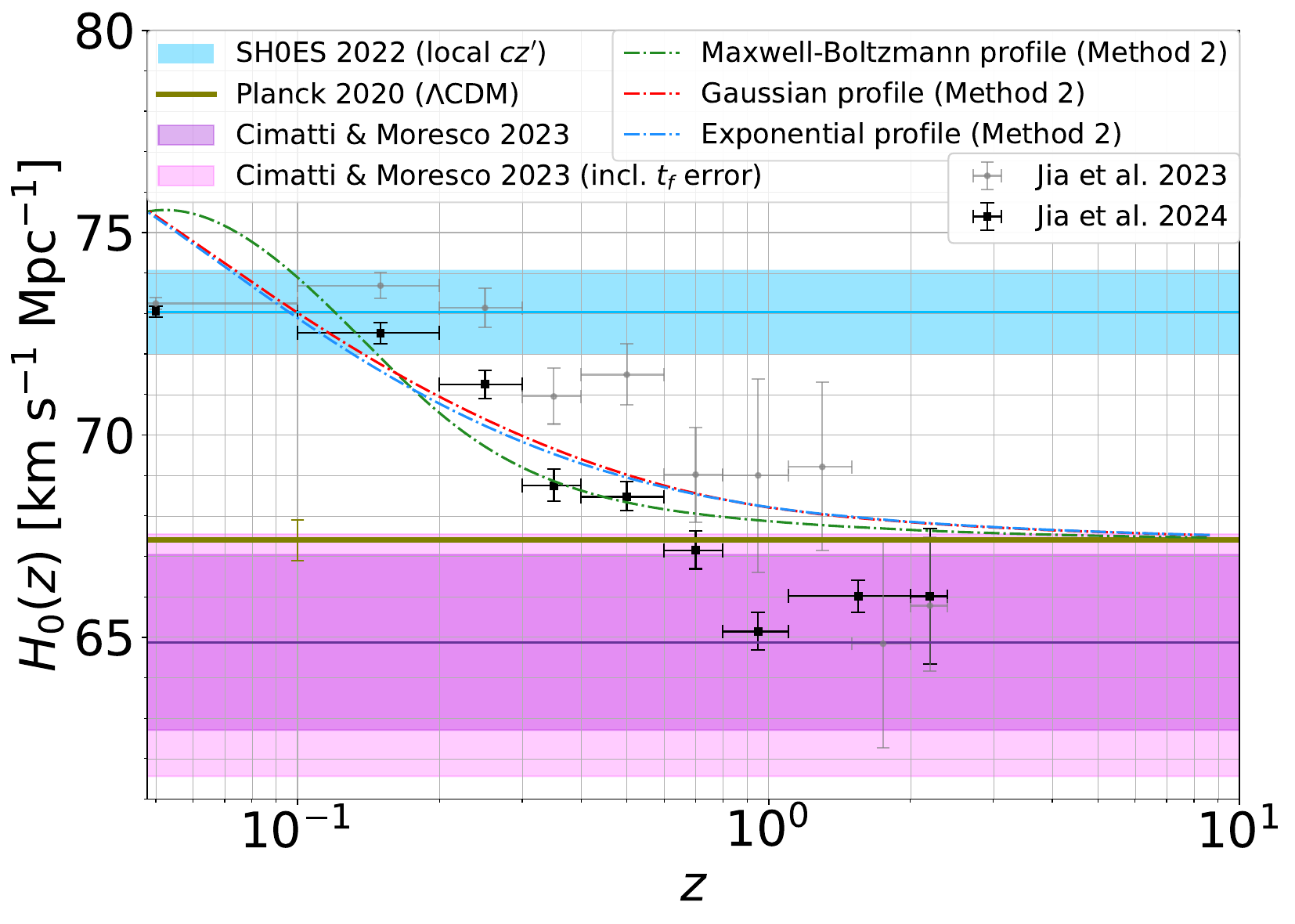}
    \caption{Similar to Figure~\ref{fig:Hubble_Diagram_Method_3_Jia_plus_noGR}, but using instead Method~1 (\emph{left panel}) and Method~2 (\emph{right panel}) for all void density profiles.}
    \label{fig:H_z_over_z_methods_1_2}
\end{figure*}

Our main results focus on $H_0(z)$ curves obtained using Method~3 (Section~\ref{sec:Method_3}) as this is most comparable to the method used by \citetalias{Jia_2023}. For completeness, we also show results using Methods 1 and 2 in the left and right panels of Figure~\ref{fig:H_z_over_z_methods_1_2}, respectively. Each panel shows the results for all three void density profiles considered by \citetalias{Haslbauer_2020}, allowing a comparison to be made between the different profiles. In each case, the void parameters are the best-fitting values as published in that study. Comparing the panels of Figure~\ref{fig:H_z_over_z_methods_1_2} shows that the particular method used to infer $H_0(z)$ has little impact on the results.

\end{appendix}

\bsp 
\label{lastpage}
\end{document}